\documentclass[12pt]{article}
\pdfoutput=1
\usepackage{epstopdf}

\usepackage{graphics,epsfig}

\def\dalemb#1#2{{\vbox{\hrule height .#2pt
        \hbox{\vrule width.#2pt height#1pt \kern#1pt
                \vrule width.#2pt}
        \hrule height.#2pt}}}

\let\a=\alpha \let\b=\beta \let\g=\gamma \let\d=\delta \let\e=\epsilon
  \let\th=\theta  
\let\l=\lambda \let\m=\mu  \let\x=\xi  %\let\r=\rho

\let\w=\omega      \let\G=\Gamma \let\D=\Delta \let\Th=\Theta 
\let\X=\Xi  \let\S=\Sigma  \let\Y=\Psi
 
\let\la=\label  
  
\def\nn{\nonumber} \def\bd{\begin{document}} \def\ed{\end{document}}
\def\ds{\documentstyle} \let\fr=\frac \let\bl=\bigl \let\br=\bigr
\let\Br=\Bigr \let\Bl=\Bigl
\let\bm=\bibitem
\let\na=\nabla
\def\tU{{\widetilde U}}
\let\pa=\partial \let\ov=\overline
\def\ie{{\it i.e.\ }}
\newcommand{\be}{\begin{equation}}
\newcommand{\ee}{\end{equation}}
\def\ba{\begin{array}}
\def\ea{\end{array}}
\def\ft#1#2{{\textstyle{{\scriptstyle #1}\over {\scriptstyle #2}}}}
\def\fft#1#2{{#1 \over #2}}
\def\F#1#2{{ F_{#1}^{(#2)} }}
\def\cF#1#2{{ {\cal F}_{#1}^{(#2)} }}

\def\R{{\bf R}}
\def\sst#1{{\scriptscriptstyle #1}}
\def\oneone{\rlap 1\mkern4mu{\rm l}}
\def\e7{E_{7(+7)}}
\def\td{\tilde}
\def\wtd{\widetilde}
\def\im{{\rm i}}
\def\bog{Bogomol'nyi\ }
\newcommand{\ho}[1]{$\, ^{#1}$}
\newcommand{\hoch}[1]{$\, ^{#1}$}
\newcommand{\bea}{\begin{eqnarray}}
\newcommand{\eea}{\end{eqnarray}}
\newcommand{\ra}{\rightarrow}
\newcommand{\lra}{\longrightarrow}
\newcommand{\Lra}{\Leftrightarrow}
\newcommand{\ap}{\alpha^\prime}
\newcommand{\bp}{\tilde \beta^\prime}
\newcommand{\cB}{{\cal B}}
\newcommand{\cO}{{\cal O}}
\newcommand{\vecx}{\vec{x}}
\newcommand{\vecy}{\vec{y}}
\newcommand{\vecp}{\vec{p}}
\newcommand{\vecq}{\vec{q}}
\newcommand{\tr}{{\rm tr} }
\newcommand{\Tr}{{\rm Tr} }
\newcommand{\NP}{Nucl. Phys. }

\newcommand{\cL}{{\cal L}}
\newcommand{\cA}{{\cal A}}
\newcommand{\cD}{{\cal D}}

\def\sst#1{{\scriptscriptstyle #1}}
\def\0{{\sst{(0)}}}
\def\1{{\sst{(1)}}}
\def\2{{\sst{(2)}}}
\def\3{{\sst{(3)}}}
\def\4{{\sst{(4)}}}
\def\5{{\sst{(5)}}}
\def\6{{\sst{(6)}}}
\def\7{{\sst{(7)}}}
\def\8{{\sst{(8)}}}
\def\ve{\varepsilon}
\def\vf{\varphi}
\def\F{\Phi}
\def\wg{\wedge}

\newcommand{\tamphys}{\it %Center for Theoretical Physics,\\
}

\newcommand{\auth}{AUTHORS}

\def\thb{\bar{\theta}}
\def\Thb{\bar{\Theta}}
\def\barp{\bar{p}}
\def\barq{\bar{q}}
\def\barc{\bar{c}}
\def\bard{\bar{d}}
\def\e{\epsilon}

\def \bi{\bibitem}
\def \la {\label}

\def \l {\lambda}
\def\foot{\footnote}
\def \tl  {{\tilde \l}}
\def \sql {{\sqrt \l}}
\def \adss {$AdS_5 \times S^5$\ }
\newcommand{\rf}[1]{(\ref{#1})}
\def \ov {\over}

\def\th{\theta}
\def\Th{\Theta}
\def\vth{\vartheta}
\def\btheta{{\bar\theta}}
\def\ttheta{{{\tilde\theta}}}
\def\bttheta{{{\bar\ttheta}}}
\def\vth{\vartheta}

\def\ra{\rightarrow}
\def\N{{\cal N}}
\def\F{{\cal F}}
\def\uM{\underline{M}}
\def\uN{\underline{N}}
\def\uP{\underline{P}}
\def\cc{\circ}
\def\eqv{\equiv}

\def\ni{\noindent}

\def\Ep{E^{{}^{(+)}}}
\def\Em{E^{{}^{(-)}}}

\def\Mp{M^{{}^{(+)}}}
\def\Mm{M^{{}^{(-)}}}

\def \ha{{1\ov 2}}

\def\r{\rho}

\def\Y{{\rm Y}}
\def\X{{\rm X}}
\def\tY{\tilde{\rm Y}}
\def\tX{\tilde{\rm X}}
\def\dY{\dot{\rm Y}}
\def\dX{\dot{\rm X}}

\def \J {\mathcal{J}}
\def \del {\partial}

\def\dF{\dot{F}}
\def\dG{\dot{G}}
\def\df{\dot{f}}
\def \E {{\cal E}}
\def \S {{\cal S}}
\def \J {{\cal J}}

\def\ms{\mathcal{S}}
\def\mj{\mathcal{J}}
\def\soj{\fr{\ms}{\mj}}
\def \R {{\bf R}}
\def \om {\omega}
\def \bE {\bar E}
\def \x {{\cal X}}

\def \bi{\bibitem}
\def \la {\label}

\def \l {\lambda}
\def\foot{\footnote}
\def \tl  {{\tilde \l}}
\def \sql {{\sqrt \l}}
\def \adss {$AdS_5 \times S^5$\ }
\def \ov {\over}

\def \varpi {{\rm w}}

\def\thb{\bar{\theta}}
\def\Thb{\bar{\Theta}}
\def\zb{\bar{z}}
\def\psib{\bar{\psi}}
\def\barp{\bar{p}}
\def\barq{\bar{q}}
\def\barc{\bar{c}}
\def\bard{\bar{d}}
\def\e{\epsilon}
\def\wb{\bar{w}}
\def\lb{\bar{\l}}
\def\Jb{\bar{J}}
\def\Nb{\bar{N}}
\def\pab{\bar{\pa}}

\def\At{\tilde{A}}
\def\Bt{\tilde{B}}
\def\Ct{\tilde{C}}
\def\Dt{\tilde{D}}
\def\Et{\tilde{E}}
\def\Ft{\tilde{F}}
\def\Gt{\tilde{G}}
\def\Mt{\tilde{M}}
\def\at{\tilde{a}}
\def\bt{\tilde{b}}
\def\ct{\tilde{c}}
\def\dt{\tilde{d}}
\def\et{\tilde{e}}
\def\ft{\tilde{f}}
\def\gt{\tilde{g}}
\def\ola{\overleftarrow}
\def\ora{\overrightarrow}
\def\alt{\tilde{\a}}

\def\eh{\hat{e}}
\def\eph{\hat{\e}}
\def\ph{\hat{p}}
\def\alh{\hat{\a}}
\def\beh{\hat{\b}}
\def\gah{\hat{\g}}
\def\muh{\hat{\m}}
\def\thh{\hat{\th}}
\def\dh{\hat{d}}
\def\deh{\hat{\d}}
\def\wh{\hat{w}}
\def\lah{\hat{\l}}
\def\Ch{\hat{C}}
\def\Omh{\hat{\Omega}}

\def\ps{\rlap{\, /}\;\,p }
\def\ks{\rlap{\, /}\;\,k }

\def\gym{g_{YM}}

\def\adot{\dot{a}}
\def\bdot{\dot{b}}
\def\bpa{\bar{\pa}}

\begin{document}
\overfullrule=0pt
\parskip=2pt
\parindent=12pt
\headheight=0in \headsep=0in \topmargin=0in
\oddsidemargin=0in

\vspace{ -3cm}
\thispagestyle{empty}
%\vspace{1cm}
%\begin{flushright}
%Preprint DFPD 01/TH/\\
%hep-th/
%\end{flushright}

 \vspace{0.1cm}

\setcounter{equation}{0}
\setcounter{footnote}{0}
\setcounter{section}{0}

\begin{center}

{\Large\bf  SYM anomalous dimension from open string engineered curvature
 %\\
 %\vspace{0.1cm}
  }

 \vspace{.5cm} { I.Y. Park%\footnote{Permanent address:
 % Philander Smith College,
%Little Rock, AR 72223, USA
 % }
 }
 \vskip 0.2cm

%%\vspace{0.8cm}
%{\it Kavli Institute for Theoretical Physics,\\
% Santa Barbara, California, USA \\
%} \vspace{0.3cm}

%\vspace{0.5cm}
%{\it Korea Institute for Advanced Study\\
%Seoul 130-012, Korea \\
%}
% \vspace{0.3cm}

\vspace{0.5cm}
{\it Department of Natural and physical Sciences,
Philander Smith College%\footnote{Home institute}
                               \\
Little Rock, AR 72223, USA \\
inyongpark05@gmail.com}

%\vspace{0.5cm}
%{\it Department of Physics and Research Institute of Basic Sciences\\
%Kyung Hee University\\
%Seoul 130-701, Korea \\
%}

\end{center}

 \vspace{0.1cm}

 \begin{abstract}
 %%%%%%%%%%%%%%%%%%%%%%%%%%%%%%%%%
In the full type IIB open superstring setup, we show that there exists a
vertex operator, denoted by $V_L$, whose renormalization reproduces
the same anomalous dimension matrix that was obtained by Minahan and Zarembo (MZ)
in {\em JHEP} {\bf 0303}, 013 (2003). The vertex operator is a gauge
singlet, and has the same SO(6) index structure as the operator used by MZ.
To obtain the SYM anomalous dimensions, one of the key ingredients
is the non-abelianization of the non-linear sigma model action in a D-brane background.
 In {\em Phys.\ Lett.} {\bf B 660}, 583 (2008), it was conjectured that quantum effects of open
strings on a stack of D-branes should generate the curvature. Even though the conjecture had passed
a few consistency checks, the role of the higher order curvature terms has remain mostly unobserved
until now. By choosing a special form of the non-abelianized counter vertex, $V_G$, we show that
the SYM result is reproduced. We comment on fermionic completion of $V_L$ and on how the special form of the non-abelian vertex $V_G$ may arise.

\end{abstract}
\newpage

\setcounter{equation}{0}
\setcounter{footnote}{0}
\setcounter{section}{0}

%\renewcommand{\theequation}{1.\arabic{equation}}
% \setcounter{equation}{0}

%%%%%%%%%%%%%%%%%%%%%%%%%%%%%%%%%%%%%%%%%%%%%%%%%%%%%%%%%%%%%%%%
\section{Introduction}
%%%%%%%%%%%%%%%%%%%%%%%%%%%%%%%%%%%%%%%%%%%%%%%%%%%%%%%%%%%%%%%%

An open string undergoes a transformation to a closed string when the
endpoints become confined \cite{Nielsen:1973qs}\cite{Gibbons:2000hf}\cite{Sen:2000kd}. The
transformation is a drastic change of the {\em degrees of freedom}, and does not have an
analogue in point particle theory. What was argued in \cite{Park:2001bm} was that the endpoints
stick together in a strong coupling limit that can be reached by open string S-duality.
(In a low energy, open string S-duality may be realized as a superspace Legendre transformation (see, e.g.,\cite{GonzalezRey:1998uh}).)
This phenomenon has an intriguing implication in the context of D-brane physics
\cite{Polchinski:1995mt}\cite{pol}. Before the conversion of an open string to a closed string,
the dynamics of the hosting D-brane is described through the Dirichlet boundary conditions. How would
it be described after the conversion?: it should be described as a soliton solution of closed string
theory (hence the title of \cite{Park:1999xz}). Open string quantum effects become increasingly important
as the coupling becomes stronger. Eventually the closed string degrees of freedom would emerge in
the strong coupling limit. Could this suggest that open string quantum effects would reveal some
closed string attributes (in a way that has not been understood before)?

The answer should be affirmative, and it was proposed in
\cite{Park:2007mc}\cite{Park:2008sg} that the quantum effects should
build themselves towards the curvature of the D-brane
geometry. The geometry would be full D-brane geometry but
not that of AdS. The latter will arise once S-duality is applied in
the open string setup. We believe that AdS/CFT can be derived along the
line of this logic. We will comment further on this in the
conclusion.

To test the validity of the proposal, a setup drawn in
Fig.1 was used: scattering at loop orders (a one-loop four-point
amplitude shown) among the open strings attached to a stack of the
"system" D-branes. The physical picture proposed depicts the
curvature around the system branes as generated by quantum effects of
the resident open strings. At the technical level, it was
anticipated that the curvature would enter through the counter
vertex operators that need to be added to cancel the divergences of
the loop diagrams. To some extent, the proposed mechanism is
reminiscent of the Fischler-Susskind mechanism \cite{Fischler:1986tb}
\cite{Fischler:1986ci}\cite{Das:1986dy}. We refer to the last paragraph of
\cite{Park:2008fp} for more detailed account of the relation of our proposal with the Fischler-Susskind mechanism. In the actual
technical setup, a parameter, $r_0$, was introduced. The physical
meaning of the parameter should be that it signals the presence of a
brane,\footnote{This point has benefited from discussions with T.
Suyama and J. Polchinski among others.} which should be
"virtual"\footnote{We call it "virtual" because its presence is
manifested only through the parameter $r_0$, and we do not introduce
a sector of open strings with one end on the system branes and the
other on the virtual brane. It should be noted that the word "virtual"
does not indicate virtual states that go around a loop in quantum field theory.},
and parameterizes the "distance" between
the system branes and the virtual branes.

The original motivation for introduction
of such counter terms was to incorporate the curvature
effect (that is produced by the D-brane) in the open string physics.
A D-brane is described through the open string boundary conditions
as a {\em flat} hyperplane. One can also consider an open string
propagating in a curved background in the usual sense. It is just
that the curved background happens to be that of a D-brane. What bridges the two, the open string attached to
the flat hyperplane and an open string propagating in the curved
background? What has been conjectured in
\cite{Park:2007mc}\cite{Park:2008sg} is that the curvature should arise from
quantum effects of the open string. If one considers the curved
background from the beginning, it would correspond to a "top-down"
approach. In a "bottom-up" approach: one would start with the flat
hyperplane, and carry out the "renormalization" program; the open
string would become finite in the {\em curved} background. It is
useful to keep both these approaches in mind.

\begin{figure}
\centerline{
\begin{minipage}[b]{9cm}
             \epsfxsize=10cm
              \epsfbox{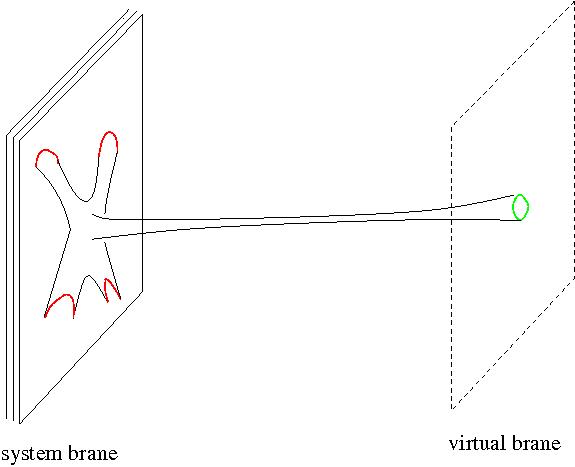}
      \end{minipage}
      }
\caption{four-point one-loop scattering}
\label{loopfig}
\end{figure}

\ni AdS/CFT has gained an additional door for exploration after the authors of \cite{Gubser:2002tv} has initiated the study of {\em solitonic configurations} of a closed string on $AdS_5\times S^5$.
Expanding the work of \cite{Gubser:2002tv} in the SYM side, Minahan and Zarembo \cite{Minahan:2002ve} considered a cyclically symmetric operator,
%%%
 \bea
 {\cal O}[\psi]=\psi^{m_1,...,m_L}\,\tr\, \phi_{m_1}\cdots \phi_{m_L}
 \label{o-quoted}
 \eea
%%%
where $\phi_{m}$'s are the scalar fields with $m$ the SO(6) index
and $\psi^{m_1,...,m_L}$ is a numerical tensor that enforces cyclic
symmetry. By analyzing the divergence structures of the self-energy
diagram and the one-loop planar interactions between two given legs,
they obtained an anomalous dimension matrix, and noticed that the
matrix takes a form of the Hamiltonian of an integrable spin chain.
The Bethe ansatz techniques were then employed to find the
eigenvalues, which are the anomalous dimensions of the operator
\rf{o-quoted}. It led to many successful comparisons between the SYM
anomalous dimensions and the energies of the corresponding closed
string configurations. In this work, we explore the possibility of obtaining the SYM anomalous dimension matrix in an open string setup.

The status of the proposal of \cite{Park:2007mc}\cite{Park:2008sg} is as follows.
In \cite{Park:2008fp}, one-loop check was carried out. The two-loop
has been partially checked in \cite{Park:2009ki}. For the
results so far, it is only the quadratic terms in the
action that played a role to cancel the divergence.  The
scattering of massive string, undertaken in \cite{arXiv:1101.1204} in the current context, may have a possibility that the higher curvature terms
contribute at one- or two- loop order. (Recall it seems to take the
three-loop order in the massless case.) The present work is in the same spirit as \cite{arXiv:1101.1204}
for the operator to be analyzed will be "heavy".
We report our observation that there exists a
vertex operator $V_L$, constructed out of open string fields, whose renormalization reproduces
the same anomalous dimension matrix that was obtained by Minahan and Zarembo.
The operator is a gauge singlet, and has the same SO(6) index structure as the operator used by MZ. It is not clear for the reasons that will be explained whether $V_L$ could be viewed as the open string analogue of \rf{o-quoted}
in the usual sense. Regardless, it has certain desired and/or necessary conditions to potentially represent a closed string state. We carry out renormalization
of $V_L$; to our surprise, the sought-for role of the higher order terms of the geometry becomes very evident even at one-loop level as will be demonstrated below.
\\

\ni The rest of the paper is organized as follows.
In the next section, we review the strategy. We point out among other things
the need for new Chan-Paton type rules for the vertices ($V_L$, $V_G$) and the relevance of world-sheet renormalization in addition to the space-time renormalization.
In section 3, we propose the vertex operator $V_L$ largely
based on the index structures of \rf{o-quoted}: having SO(6)- and gauge-
structures that are parallel to those of \rf{o-quoted}, $V_L$ reproduces the SYM result in the low energy limit.
 Although the index structures of $V_L$ are suggestive, it is not clear
whether the vertex operator $V_L$ could be identified as the open string
theory analogue of \rf{o-quoted} in the usual sense.
Some subtle issues must be understood to see whether such relationship exists.
 We use Green-Schwarz formulation because the
tasks just mentioned and computations can be carried out most easily. In section 4, We carry out the spacetime and worldsheet
renormalization, and determine the renormalization constant of $V_L$ after taking the "small $\a'$"
limit. We mostly focus on $L=2$ case for explicitness, which captures many of the essential
features of the analysis. We show that the index structure of the constant is exactly the same
as the SYM case. $L\geq 3$ cases will be commented on afterwards. A more comprehensive account
including details and various reviews will be given elsewhere \cite{ipark}. We conclude with remarks and future directions in section 5.

%%%%%%%%%%%%%%%%%%%%%%%%%%%%%%%%%%%%%%%%%%%%%%%%%%%%%%%%%%%%%%%%
\section{Overview of strategy}
%%%%%%%%%%%%%%%%%%%%%%%%%%%%%%%%%%%%%%%%%%%%%%%%%%%%%%%%%%%%%%%%

Compared with the SYM analysis, there are some differences in
the stringy renormalization of $V_L$ as we will see below. One very drastic
difference is that the one-loop string correlator
%%%
 \bea
<V_s(x_1)V_s(x_2)V_s(x_3)\cdots V_s(x_L);V_{L}(x)\;>_{one-loop}
\label{oneloop_by Green}
 \eea
 %%%
 vanishes unlike the SYM case. The vertex operator $V_s$ represents scalar state \cite{Park:1999xz}, and is defined by
%%%
 \bea
  V_s\equiv ({X'}^{m}+R^{mv}k^{v})\;e^{ik\cdot X}
    \label{scalarvertex}
 \eea
 %%%
 where $m$ is the SO(6) index and $k^v$ is a momentum along the brane worldvolume. $R^{mv}$ is the fermionic bilinear,
 %%%
 \bea
 R^{mv}\equiv \fr14 S\g^{mv} S
 \eea
 %%%
 where the $S$ is a re-scaled coordinate of the fermionic coordinate, $\th$ \cite{gsw}.
We refer to, e.g.,\cite{Park:2007mc}\cite{Park:2008sg} for more details of our notations.
As a matter of fact, this is a very telling difference since it implies that only the
contributions from counter vertices remain relevant. We will see that the role of the quartic order vertex
that comes from $V_G$
is essential: otherwise the SYM result cannot be reproduced. Therefore the present analysis
explicitly demonstrates the role of the higher order curvature terms.

Because of this difference and other subtle issues, it may be useful to have an overview
on the strategy in general before we get to the detailed analysis.
One of the crucial issues is {\em worldsheet} renormalization. This issue was not
dealt with in our previous works where we focused on spacetime renormalization. After
introducing the spacetime counter vertices, one should make sure that the worldsheet results
are made finite.
Let us elaborate on the idea. To keep the discussion heuristic, we will suppress
the gauge indices.
The {\em spacetime} renormalization was conjectured \cite{Park:2007mc}\cite{Park:2008sg} to
introduce the geometric counter vertex, $V_G$.
 We focus on the following part of $V_G$
 %%%
 \bea
 &&-\fr{q}{2}\int \;h^{st}\;
     \pa_s X^m \pa_t X^m\left(  \frac{1}{2}-\fr{2X_0\cdot X}{r_0^2}-\frac{1}{{r_0}^2}
                X^nX^n+\fr{6(X_0\cdot X)^2}{r_0^4}\right)
                \label{bctr}
 \eea
 %%%
 where $h^{st}$ is the flat worldsheet metric and the parameter $q$ is defined in terms of the open string constant, $g$, by
 %%%
 \bea
  q\equiv \frac{4\pi g^2\a'^2}{r_0^4}
 \eea
 %%%
  The vector $X_0^n$ specifies the location of the virtual brane. The fermionic bilinear
term in $V_s$ will not play a role: it will lead to a disconnected diagram at most, and
therefore will not require separate consideration. (Strictly speaking, this will be the case when one uses $V_L$ given in \rf{VLna}.
We will comment on this further later.) The reason is that it will have be contracted
with another bilinear or $V_G$ but not with $V_L$ since in the lightcone gauge $V_L$ does not
contain any fermionic field, at least in the zero momentum case.
 Since the external vertices do not contain any fermionic fields, the fermionic part of $V_G$ can
be entirely omitted in dimensional regularization. In any case, the fermionic part has
a stronger tendency to lead to vanishing results than the bosonic vertices as observed in the
previous works \cite{Park:2008fp}\cite{Park:2009ki}.

\ni With the vertex operator $V_{L}$ is determined in the next section, one should compute
correlators at tree and
one-loop level:
 %%%
 \bea
<V_s(x_1)V_s(x_2)V_s(x_3)\cdots V_s(x_L);V_{L}(x)>_{tree}
 \label{general_tree}
 \eea
 %%%
 %%%
 \bea
<V_s(x_1)V_s(x_2)V_s(x_3)\cdots V_s(x_L);V_{L}(x)\;>_{one-loop}
\label{oneloop_by Green}
 \eea
 %%%
 and the third correlator with the counter vertex inserted:
 %%%
 \bea
<V_G(y) V_s(x_1)V_s(x_2)V_s(x_3)\cdots V_s(x_L);V_{L}(x)\;>_{tree}
\label{oneloop_ctr}
 \eea
 %%%
 For correlators \rf{general_tree} and \rf{oneloop_ctr}, one uses the tree-level Green's function as indicated by
"tree". The result is of the same order of the coupling constant as
\rf{oneloop_by Green} because of the coupling constant dependence of
$V_G$. The tree level computation is straightforward although
tedious. In the SYM case, there are three different one-loop
diagrams as given in Fig 2. For the string analysis, the spacetime
one-loop vanishes due to the impossibility of saturating the zero
modes. The {\em worldsheet analysis} parallels the SYM analysis to
some extent but there exists subtle differences. First of all, there
is no diagram that correspond to Fig. 2(a). There are two worldsheet
diagrams that respectively correspond to the last two diagrams in
Fig. 2, and they are both at at the order of $q$. The
renormalization constant for $V_L$ is determined in such a way that
with \rf{oneloop_ctr} it cancels the one-loop divergence of
\rf{oneloop_by Green}.\footnote{We do not consider spacetime
renormalization of the string tension $T$ for the current work in order to keep the
procedure at minimal level. } But since the spaetime one-loop
vanishes, it is fixed so as to absorb the divergence of
\rf{oneloop_ctr}.

\begin{figure}
\centerline{
\begin{minipage}[b]{9cm}
             \epsfxsize=10cm
              \epsfbox{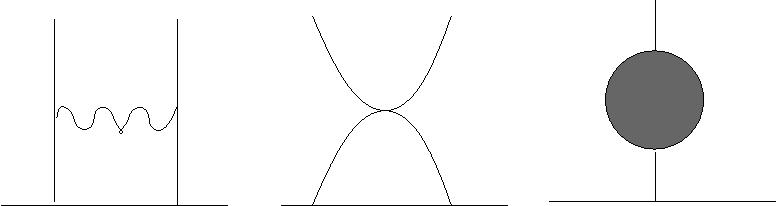}
      \end{minipage}
      }
\caption{one-loop diagrams in SYM}
\label{MZfig}
\end{figure}

%%%%%%%%%%%%%%%%%%%%%%%%%%%%%%%%%%%%%%%%%%%%%%%%%%%%%%%%%%%%%%%%
\section{Proposal of $V_L$}
%%%%%%%%%%%%%%%%%%%%%%%%%%%%%%%%%%%%%%%%%%%%%%%%%%%%%%%%%%%%%%%%

In carrying out renormalization\footnote{A fully proper analysis would require off-shell
computations and, therefore, only be possible in a string field theory setup. We will come back to this issue
below.} of $V_L$
the SYM analysis of \cite{Minahan:2002ve}, there are subtle and crucial issues that
one expects to face. One of them is associated with the Chan-Paton method.
   Let us consider the sigma model in the background of the D-brane geometry.
More specifically, the large-$r_0$ expanded, the non-linear sigma model contains a quartic term such as
 %%%
 \bea
  \pa X^m \pa X^m X^nX^n \label{quartic vertex}
 \eea
 %%%
 In the single brane case, one may take the top-down approach and use the vertex above
as a counter term. What about the case of a multi-brane case? As for the scattering of the
elementary string states, the multiplicity of the system branes is taken into account by
standard Chan-Paton factors. The question is how to incorporate the Chan-Paton factors
for a type of a composite vertex such as \rf{quartic vertex}. Once we postulate the presence
of such terms, the need for their non-abelianization seems evident: it would be unnatural to
keep the same abelian forms of the counter vertices while the elementary states and $V_L$ are
treated non-abelianized through the usual Chan-Paton method.
\\

Let us turn to the proposed form of the open string vertex $V_L$.
Currently there are three superstring formulations available: Ramond-Nevu-Schwarz,
Green-Schwarz and pure spinor \cite{Berkovits:2000fe}.
 Each formulation has its advantages and disadvantages; we employ Green-Schwarz
formulation in this work.

Let us imagine constructing the vertex operator $V_L$ basically by
multplying the open string creation
operators. This immediately poses several subtle issues. The first is the question whether one can take a product of a {\em finite} number of creation operators as against an infinite number. In other words, could a vertex operator constructed out of
a product of a finite number of the open string coordinates represnt
a closed string state? It is believed (see, e.g.,\cite{Print-87-0006}) that a closed string state does not belong to the open string Fock space.\footnote{Following \cite{Print-87-0006}, a Fock state is defined to be a finite linear combination of states with a finite number of creration opearators acting on each state.} The geometric vertex
operator $V_G $ is indeed an infinite series of the open string
coordinates once it is large-$r_0$ expanded \cite{Park:2008fp}, hence not in the open string Fock space. (In this work, we only consider the relevant parts of $V_G$ suppressing its infinte tail, but that is a matter of notational convenience. In other words, the omitted parts are irrelevant for the given orders
of the perturbation theory.)
We will pursue the resolution of this issue elsewhere but instead
contemplate below on a feature
of $V_L$ that is another necessary condition for $V_L$ to represnt a closed string state.
($V_G$ shares the same feature.)

Secondly, there is a question on the physical state condition.
In the first quantized string, the physical state condition is tied with conformal symmetry, and it
leads to an on-shell momentum of the vertex operator.
However, it is unclear whether the physical
state condition must be imposed on a vertex operator that
is supposed to represent a closed string. It is especially
so if, indeed, it is required to take $L=\infty$ to
properly represent a closed string since the vertex then would
lie outside the Fock space. We will set these issues aside for futher
discussion in the conclusion.

In the next section, we consider
  %%%
 \bea
V_{L}\equiv  \psi_{m_1...m_L}\,{X'}^{m_1}\cdots {X'}^{m_L}\;
 \label{VL}
 \eea
 %%%
carrying out its renormalization, and note that it leads to the same
anomalous dimension matrix as the SYM case.
More precisely, it is its non-abelian version of \rf{VL} that will be analysed,
%%%
 \bea
V_{L}\equiv  \psi_{m_1...m_L}\,[{X'}^{m_1}\otimes a_1]\cdots [{X'}^{m_L}\otimes a_L]\;
   \Tr(\l^{a_1}\cdots \l^{a_L})
 \label{VLnaq}
 \eea
 %%%
which is eq.\rf{VLna} in section 4.
  We will do that without full justification on its form; it is mostly the index structure  of \rf{o-quoted} and the operator-state mapping that led us to consider \rf{VLnaq}.
Even though we do not have full justification, there are several
practical reasons for considering \rf{VLnaq}. Simply put, we believe that
\rf{VLnaq} could very well be the bosonic part of the properly formulated full vertex
operator. The full form of the operator may contain the
fermionic coordinates as well. However, the terms with the fermionic
coordinates tend to lead to vanishing results as observed in e.g.,
\cite{Park:2009ki}.
 More detailed remarks will
be given in the conclusion. Note that the vertex above has a gauge structure
that is very different from, say, that of an ordinary massive
open string; in particular, it is
a gauge singlet. It is in contrast with the fact that an ordinary open string has uncontracted adjoint indices prescribed by the usual Chan-Paton method. A vertex operator should be a gauge singlet
in order to represent a closed string. (The same applies to the non-abelian form of the geometric vertex $V_G$ as we will see shortly.)

%%%%%%%%%%%%%%%%%%%%%%%%%%%%%%%%%%%%%%%%%%%%%%%%%%%%%%%%%%%%%%%%
\section{Anomalous dimension of $V_L$}
%%%%%%%%%%%%%%%%%%%%%%%%%%%%%%%%%%%%%%%%%%%%%%%%%%%%%%%%%%%%%%%%

The renormalization that was carried out in the previous works
\cite{Park:2008fp}\cite{Park:2009ki} was spacetime renormalization.
It (conjecturally) introduces higher order composite operators. Once
such operators are present, one should ensure that the worldsheet
results are made finite. In other words, one should carry out
worldsheet renormalization in addition to the spacetime
renormalization. As shown below, it is a neat interplay between the
worldsheet renormalization and the spacetime renorma1ization that
reproduces the SYM result.

 The renormalization is different from that of an unrenormalizable field theory where one does not have a control over proliferation of counter terms. The forms of the counter terms are already fixed by the supergravity solution other than a few parameters such as the string tension. As in any regular quantum field theory,
renormalization of elementary field must be carried prior to renormailzation of a composite operator.

 The conceptual order of the
renormalization program  is as follows. We accept as true the
conjecture of \cite{Park:2007mc} (or simply postulate) that the
spacetime renormalization introduces
 the higher order curvature terms. The worldsheet renormalization program is carried out introducing worldsheet counter terms. In particular, the wave function will become worldsheet-renormalized. Up to this point, the vertex operator $V_L$ is irrelevant.
 Consider the spacetime quantum corrections. For the present work, we consider one-loop. Then carry out the worldsheet renormalization inserting $V_L$.
  The spacetime one-loop amplitudes vanish as we will show below. Therefore, only the contributions from the geometric counter vertices remain. Finally, determine the renormalization constant of $V_L$ while ensuring the finiteness of correlators. By fixing the relative coefficients appropriately between spacetime counter terms and the worldsheet counter terms, one arrives at the SYM result after taking the "small $\a'$" limit.

 As we will see,
the parameter $\a'$ can be scaled away in the large-$r_0$ expansion. Because of this,
the higher worldsheet loop expansion is no longer parametrized by $\a'$, but  a new set of
parameters come to serve as the parameters for the worldsheet loop expansion. This implies that
the would-be small $\a'$ limit is replaced by a limit of these parameters.

With the preliminary discussion above, we are ready for more detailed analysis of the worldsheet renormalization.
Introduction of \rf{bctr}, in turn, necessitates worldsheet renormalization. In particular, the cubic vertex leads to wave function renormalization as shown in Fig.3. (The spacetime one-loop does not lead to field renormalization: it is a worldsheet effect. (Since it is $\pa X \pa X$ it is not a mass renormalization.)) This implies among other things that the quadratic vertex\footnote{The one-loop diagram in Fig.3 comes at $q^2$ order.  Although $V_{q^2}$ has $\pa X \pa X$ vertex which will get renormalized because of Fig.3, it is irrelevant since it will have to come with $\fr1{r_0^4}$.} gets renormalized, and that is essential for reproduction of the SYM result.

\begin{figure}
\centerline{
\begin{minipage}[b]{7cm}
             \epsfxsize=7cm
             \epsfbox{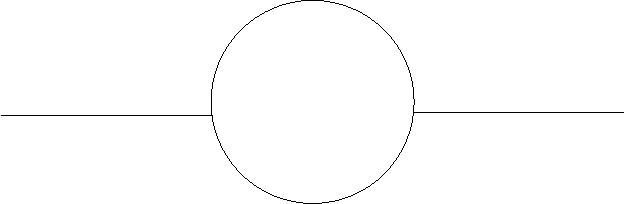}
      \end{minipage}
      }
\caption{worldsheet wave-function renormalization}
\label{wsren}
\end{figure}

\ni The abelian case is fairly easy to carry out and details will be presented in \cite{ipark}. Here we summarize the result and comment on the non-abelian counter part. The upshot of the analysis is that
with the vertex of \rf{bctr} present, the following combination will be relevant for the renormalization of $V_L$,
 %%%
 \bea
   \fr{\G(1-\w)}{r_0^2}
   \frac{h^{st}}{2}\pa_s X^m \pa_t X^m -\frac{1}{{r_0}^2}
               h^{st} \pa_s X^m \pa_t X^m X^nX^n
               \label{renctr}
 \eea
 %%%
 where the first term comes from the worldsheet wave function renormalization. 2D
 dimensional regularization was used: the parameter $\w$ is the usual parameter in dimensional regularization with $\w\ra 1$.
 The second term is the third vertex in \rf{bctr}. A few remarks are in order: The terms \rf{bctr} that contain $X_0$ lead to vanishing correlators due to SO(6) invariance, and do not contribute to renormalization of $V_L$. The fourth term is in the higher order of the "small $\a'$" limit - which we will discuss shortly - and can be omitted. The quartic vertex produces terms in higher orders as well but they have been omitted for the same reason. We now comment on the non-abelian generalization whose fuller account is given below. We propose that the internal vertices (i.e., the counter vertices) should be non-abelianized by "taking a tensor product" with the group indices in a certain manner. In particular, it will modify the first term of \rf{renctr}:
 %%%
 \bea
   N\,\fr{\G(1-\w)}{r_0^2}\;
   \frac{h^{st}}{2}\pa_s X^m \pa_t X^m -\frac{1}{{r_0}^2}\,
               h^{st}\, \pa_s X^m \pa_t X^m X^nX^n
               \label{renctr-nonabelian}
 \eea
 %%%
 where we have suppressed the tensor product with the group generators that was mentioned above.
A more explicit expression will be presented below.
The appearance of the extra factor $N$ for that particular vertex has been deduced on general grounds but not rigorously derived here. We will come back to this after we give a fuller account of the non-abelianization. The "small $\a'$" limit or the limit that one should take for comparison with the SYM result is basically $r_0\ra \infty$ limit. Also since the SYM result is obtained in the planar limit, one should take large $N$ limit in such a way
 %%%
 \bea
 \frac{N}{r_0^2}\;\;\; \mbox{is small and finite.} \label{sym limit}
 \eea
 %%%
 We can view $r_0$ as dimensionless by rescaling the coordinates.
Now one computes \rf{oneloop_ctr} with $V_G$ given by \rf{renctr-nonabelian}. The 2D integral of the correlator will produce a finite result for the quadratic term but there is already an overall diverging coefficient in front, the $\G(1-\w)$ factor. For the quartic vertex, however, it produces an diverging result, namely $\G(1-\w)$-factor. Finally we follow the standard method, taking a derivative with respect to $\G(1-\w)$, to get the anomalous dimension matrix.\footnote{Even though the original quadratic vertex is there, its contribution drops out after the derivative.}

\vspace*{.1in}

We now ponder on the issue of non-abelianizing the counter vertices.
 The non-linear sigma model actions was originally constructed by conducting the $\beta$-function procedure in the {\em closed} string context: this action can provide the counter vertices in the case of a single brane. For the case of a multi-brane system, things are more subtle.  One can assign the standard Chan-Paton factors to external vertices such as $V_s$ or $V_L$.
 A new procedure seems necessary for the internal vertices. Combining the quadratic and quartic counter vertices, $V_G$ in the abelian case is given by
%%%
 \bea
     h^{st}\,  \pa_s X^m \pa_t X^m \oplus
               h^{st}\, \pa_s X^m \pa_t X^m X^nX^n
 \label{vqmother_rel2q}
 \eea
 %%%
 where $\oplus$ indicates that the coefficients are not recorded precisely.\footnote{When non-abelianzing these vertices in the way that we propose, the relative numerical coefficients cannot be precisely determined. It is a limitation of our proposal. For precise determination of relative coefficients in general, one would need more systematic method that is commented in the next footnote and in the conclusion. }
 How should these vertices be non-abelianized?

\ni Comparisons made with SYM must be done with care. While the SYM analysis is capable of dealing with the Green's functions, the string theory formulation can produce only scattering amplitudes. The  external scalar field in the single-line approach - and we use the single-line approach in our review of in \cite{ipark} - is $\phi^a$. (Of course it is $\phi_{ij}=\phi^a \l^a_{ij}$ in the double-line approach where $\l$ is a group generator.) In the string theory framework, the two bases are related by
 %%%
 \bea
 |k;a>=\sum |k;ij>\l^a_{ij} \label{twobases}
 \eea
 %%%
In the string scattering formulation, a vertex operator that corresponds to $|k;ij>$ is used. What is required (or more convenient) for comparison with SYM analysis is the usage of the vertex operators that correspond to $|k;a>$.
To implement this and also to maximally parallel the open string analysis with the SYM analysis, we introduce a notation where the state in the left-hand side of \rf{twobases} is represented by the following vertex operator,
 %%%
 \bea
 X'  \otimes a \label{xl}
 \eea
 %%%
When two operators $X'  \otimes a$ and $X'  \otimes b$ get contracted,
we will assume that the group parts yield  $\d^{ab}$. At this point, we take the rule as a working assumption even though it might be possible to better motivate. Proper formulation may require the use of some generalized algebra, which would also make possible the precise determination of the relative coefficient mentioned in the previous footnote. We will come back to this point in the conclusion.
The non-abelian version of $V_L$ given in \rf{VL} should be
 %%%
 \bea
V_{L}\equiv  \psi_{m_1...m_L}\,[{X'}^{m_1}\otimes a_1]\cdots [{X'}^{m_L}\otimes a_L]\;
   \Tr(\l^{a_1}\cdots \l^{a_L})
 \label{VLna}
 \eea
 %%%
As for non-abelianizing the two vertices in \rf{vqmother_rel2q}, we propose the following.
 Non-abelianization of the quadratic vertex seems straightforward:
 %%%
 \bea
 \int dy [X'^{\,i'}(y)\otimes {d'}][X'^{\,i'}(y)\otimes {e'}]\,\d_{d'e'}
 \label{quadratic vertex}
 \eea
 %%%
The quartic vertex,
 %%%
 \bea
 \int dy X'^{i'}(y) X^{j'}(y) X'^{i'}(y) X^{j'}(y)
 \eea
 %%%
 takes more efforts. As a first step, we assign one group generator to each field,
 %%%
 \bea
 \int dy [X'^{\,i'}(y)\otimes {b'}] [X^{j'}(y)\otimes {c'}] [X'^{i'}(y)\otimes {d'}] [X^{j'}(y)\otimes {e'}] \label{quartic indices}
 \eea
 %%%
  Since the final expression should be gauge invariant, the group indices above must be contracted.
   It seems that there are two choices a priori: either with Kronecker deltas or with the group structure constants. Only the latter leads to reproduction of the SYM result.\footnote{It may appear that we are intruducing the quartic terms in an artificial way for getting the same anomalous dimension matrix as the SYM analysis. We will comment on this in the conclusion.} There are three inequivalent ways of contracting the indices $(b',c',d',e')$ with the structure constants:
 %%%
 \bea
 && f_{a'b'c'}f_{a'd'e'}[X'^{\,i'}\otimes {b'}] [X^{j'}\otimes {c'}] [X'^{\,i'}\otimes {d'}] [X^{j'}\otimes {e'}]
    ,\nn\\
 &&  f_{a'b'c'}f_{a'd'e'}[X'^{\,i'}\otimes {b'}] [X'^{\,j'}\otimes {c'}] [X^{\,i'}\otimes {d'}] [X^{j'}\otimes {e'}]
    ,\nn\\
 && f_{a'b'e'}f_{a'd'c'}[X'^{\,i'}\otimes {b'}] [X'^{\,j'}\otimes {c'}] [X^{\,i'}\otimes {d'}] [X^{j'}\otimes {e'}]
 \label{quartic vertices}
 \eea
 %%%
We compute the correlators with these vertices with equal relative weights, and show that they reproduce, when summed up, the result of the SYM analysis once the correlator with \rf{quadratic vertex} is added with an appropriate relative numerical coefficient.

\vspace*{.1in}
We are ready for quantitative analysis. We start with $L=2$ and provide details.
Even though it is the simplest case, it captures many of the essential features of the cases of generic values of $L$. After that, we discuss a few aspects of $L\geq 3$ cases that are not captured by the $L=2$ analysis.

%%%%%%%%%%%%%%%%%%%%%%%%%%%%%%%%%%%%%%%%%%%%%%%%%%%%%%%%%%%%
\subsection*{$L=2$ case}

The tree level correlator without any counter vertex inserted is
%%%
 \bea
&&<[X'^{\,j}(x_2)\otimes {b}] [X'^{k}(x_3)\otimes {c}] [X'^{i_{l+1}}(x)\otimes {f}] [X'^{i_{l+2}}(x)\otimes {g}]N\d^{fg}> \nn\\
% &&= N\d^{fg}(\d_{bf}\d_{cg}\,\d^j_{i_{l+1}}\d^k_{i_{l+2}}
%               +\d_{bg}\d_{cf}\,\d^j_{i_{l+2}}\d^k_{i_{l+1}})\,
%            (\pa_{x_2}\pa_{x}\D_{x_2x})(\pa_{x_3}\pa_{x}\D_{x_3x})   \nn\\
 &&= N \d_{bc}\,(\d^j_{i_{l+1}}\d^k_{i_{l+2}}+\d^j_{i_{l+2}}\d^k_{i_{l+1}})\,
            (\pa_{x_2}\pa_{x}\D_{x_2x})(\pa_{x_3}\pa_{x}\D_{x_3x})
            \label{L2tree}
 \eea
 %%%
where $\D_{x_1x_2}$ is the two point function
 %%%
 \bea
 <X^{m_1}(x_1)X^{m_2}(x_2)>=\eta^{m_1m_2}\,\D_{x_1x_2}
 \eea
 %%%
 where we have chosen $(x_2,x_3)$ for the locations of the two elementary states.
As pointed out before for a case of arbitrary $L$, the spacetime one-loop corrections vanish. We can see this explicitly for the first few values of $L$: to saturate the zero modes, four factors of the fermionic bilinear, $R^{mv}$, would be required, but there are not sufficiently many $R^{mv}$'s for $L=2,3$. The one-loop contribution will enter with $L\geq 4$. However, even with $L\geq 4$, the result will either be a vanishing or disconnected diagram, and therefore need not be considered further.
 The correlator with the quadratic counter vertex \rf{quadratic vertex},
%%%
 \bea
<[X'^{\,j}(x_2)\otimes {b}] [X'^{k}(x_3)\otimes {c}] [X'^{i_{l+1}}(x)\otimes {f}] [X'^{i_{l+2}}(x)\otimes {g}]N\d_{fg}\;
 [X'^{\,i'}(y)\otimes {d'}][X'^{\,i'}(y)\otimes {e'}]\,\d_{d'e'}>,
 \nn\\
 \eea
 %%%
yields
 %%%
 \[
 2\Big[
 \d^{bc}\d^{fg}\; \d^{jk}\d_{i_{l+1}i_{l+2}}\,
       (\pa_{x_2}\pa_{x_3}\D_{x_2x_3}) (\pa_{x}\pa_{y}\D_{xy})^2
  \]
 %%%
 \bea
 + 2(\d^{bf}\d^{cg}+\d^{bg}\d^{cf})
    (\d_{j_{l+1}}^j\d_{k_{l+2}}^j+\d_{j_{l+2}}^j\d_{k_{l+1}}^j)
    (\pa_{x_2}\pa_{x}\D_{x_2x})  (\pa_{x_3}\pa_{x}\D_{x_3x}) (\pa_{x}\pa_{y}\D_{xy})
 \Big]\;N\d^{fg}
 \eea
 %%%
The first term is a disconnected diagram, and will therefore be omitted: one gets
 %%%
 \bea
 4N\d^{bc}\,
    (\d_{j_{l+1}}^j\d_{k_{l+2}}^j+\d_{j_{l+2}}^j\d_{k_{l+1}}^j)
    F_0
    \label{xxfinal}
 \eea
 %%%
 where
 %%%
 \bea
 F_0\equiv \int dy\int dx\, \pa_y\pa_x \D_{yx}
   \pa_y\pa_{x_1} \D_{yx_1}\pa_{x_2}\pa_x \D_{x_2x} \label{F0}
%  = \frac{1}{(x_2-x)^2} \int_{x_1+\e}^\infty
 %      \fr1{(y-x)^2}\fr1{(y-x_1)^2}=\frac{1}{x_1^2\, \e}
 \eea
 %%%
 %%%%%%%%%%%%%%%%%%5
\begin{figure}
\centerline{
\begin{minipage}[b]{2cm}
             \epsfxsize=2cm
              \epsfbox{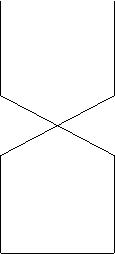}
      \end{minipage}
     }
\caption{worldsheet one-loop for $L=2$ with quartic vertex }
\label{wsren}
\end{figure}
 %%%%%%%%%%%%%%%%%%%%%%%%
\ni The correlators with the vertices \rf{quartic vertices} inserted corresponds to the diagram in Fig. 4, and lead respectively to
  %%%
 \bea
 \int dy  &&\;N\d^{fg}<
 [X'^{\,j}(x_2)\otimes\l^{b}] [X'^{k}(x_3)\otimes\l^{c}] [X'^{i_{l+1}}(x)\otimes\l^{f}] [X'^{i_{l+2}}(x)\otimes\l^{g}]\nn\\
 &&
 f_{a'b'c'}f_{a'd'e'}[X'^{\,i'}(y)\otimes\l^{b'}] [X^{j'}(y)\otimes\l^{c'}] [X'^{\,i'}(y)\otimes\l^{d'}] [X^{j'}(y)\otimes\l^{e'}]>
 \label{1stcontri}
 \eea
 %%%
  %%%
 \[
 =\;N\d^{fg}\Big[2F_1 c_1(f_{a'gf}f_{a'cb}\d_{i_{l+1}}^j \d_{i_{l+2}}^k
          + f_{a'gf}f_{a'bc}\d_{i_{l+2}}^j \d_{i_{l+1}}^k)
   + 2F_1c_1 (f_{a'gc}f_{a'bf}\d_{i_{l+2}}^j \d_{i_{l+1}}^k
          + f_{a'gb}f_{a'cf}\d_{i_{l+1}}^j \d_{i_{l+2}}^k)
  \]
 %%%
  \[
 + 2(F_2+F_3)c_1 (f_{a'gc}f_{a'fb}\d^{jk} \d_{i_{l+1}i_{l+2}}
          + f_{a'gb}f_{a'fc}\d^{jk} \d_{i_{l+1}i_{l+2}})\Big],
   %        \label{ctrxxxxonecomputed}
 \]
 %%%
  %%%
 \bea
 \int dz  &&\;N\d^{fg}<
 [X'^{\,j}(x_2)\otimes\l^{b}] [X'^{k}(x_3)\otimes\l^{c}] [X'^{i_{l+1}}(x)\otimes\l^{f}] [X'^{i_{l+2}}(x)\otimes\l^{g}]\nn\\
 &&
 f_{a'b'c'}f_{a'd'e'}[X'^{\,i'}(y)\otimes\l^{b'}] [X'^{\,j'}(y)\otimes\l^{c'}] [X^{\,i'}(y)\otimes\l^{d'}] [X^{j'}(y)\otimes\l^{e'}]>
 \label{2ndcontri}
 \eea
 %%%
  %%%
 \[
=\;N\d^{fg}\Big[2(F_2+F_3)c_2 (f_{a'fg}f_{a'cb}\d_{i_{l+2}}^j \d_{i_{l+1}}^k
          + f_{a'fg}f_{a'bc}\d_{i_{l+1}}^j \d_{i_{l+2}}^k)
   + 2F_1c_2 (f_{a'cg}f_{a'fb}\d_{i_{l+2}}^j \d_{i_{l+1}}^k
          + f_{a'bg}f_{a'fc}\d_{i_{l+1}}^j \d_{i_{l+2}}^k)               \]
 %%%
 \[
 + 2F_1c_2 (f_{a'cg}f_{a'bf}\d^{jk} \d_{i_{l+1}i_{l+2}}
          + f_{a'bg}f_{a'cf}\d^{jk} \d_{i_{l+1}i_{l+2}})\Big]
   %        \label{ctrxxxxtwocomputed}
 \]
 %%%
 and
%%%
 \bea
 \int dz  &&\;N\d^{fg}<
 [X'^{\,j}(x_2)\otimes\l^{b}] [X'^{k}(x_3)\otimes\l^{c}] [X'^{i_{l+1}}(x)\otimes\l^{f}] [X'^{i_{l+2}}(x)\otimes\l^{g}]\nn\\
 &&
 f_{a'b'e'}f_{a'd'c'}[X'^{\,i'}(y)\otimes\l^{b'}] [X'^{\,j'}(y)\otimes\l^{c'}] [X^{\,i'}(y)\otimes\l^{d'}] [X^{j'}(y)\otimes\l^{e'}]>
 \label{3rdcontri}
 \eea
 %%%
 %%%
 \[
 =\;N\d^{fg}\Big[ 2(F_2+F_3)c_3 (f_{a'cg}f_{a'fb}\d_{i_{l+2}}^j \d_{i_{l+1}}^k
          + f_{a'bg}f_{a'fc}\d_{i_{l+1}}^j \d_{i_{l+2}}^k)
   + 2F_1c_3 (f_{a'cg}f_{a'bf}+ f_{a'bg}f_{a'cf})\d^{jk} \d_{i_{l+1}i_{l+2}}
 \]
 %%%
 %%%
 \[
 + 2F_1c_3(f_{a'fg}f_{a'cb}\d_{i_{l+2}}^j \d_{i_{l+1}}^k
          + f_{a'fg}f_{a'bc}\d_{i_{l+1}}^j \d_{i_{l+2}}^k)\Big]
       %    \label{ctrxxxxthreecomputed}
 \]
 %%%
In the equations above, $F$ factors are defined as
 %%%
 \bea
 F_1 &=& \int dy\int dx\; (\pa_x\pa_y \D_{xy})(\pa_x \D_{xy}) \pa_y
             (\pa_{x_1}\D_{x_1y}\;\pa_{x_2}\D_{x_2y})\nn\\
 F_2 &=& \int dy\int dx\;(\pa_x\pa_y \D_{xy})(\pa_x\pa_y \D_{xy})
             (\pa_{x_1}\D_{x_1y})(\pa_{x_2}\D_{x_2y})\nn\\
  F_3 &=& \int dy\int dx\;(\pa_x \D_{xy})(\pa_x \D_{xy})
             (\pa_{x_1}\pa_y\D_{x_1y})(\pa_{x_2}\pa_y\D_{x_2y})
             \label{Fs}
 \eea
 %%%
 One may Fourier-transform them to isolate their divergences using dimensional regularization
  in 2D momentum space. The details will be presented in \cite{ipark}.
Adding \rf{2ndcontri}, \rf{2ndcontri} and \rf{2ndcontri} yields an expression where $(2F_1+F_2+F_3)$ factors out:
%%%
 \bea
  && 2N\d^{fg}(2F_1+(F_2+F_3))\Big[  (f_{a'fg}f_{a'bc}\d_{i_{l+1}}^j \d_{i_{l+2}}^k
      +f_{a'fg}f_{a'cb}\d_{i_{l+2}}^j \d_{i_{l+1}}^k)   \label{ctrxxxxtotoal}\\
  && + (f_{a'cg}f_{a'bf}+ f_{a'bg}f_{a'cf})\d^{jk} \d_{i_{l+1}i_{l+2}}
  + (f_{a'cg}f_{a'fb}\d_{i_{l+2}}^j \d_{i_{l+1}}^k
          + f_{a'bg}f_{a'fc}\d_{i_{l+1}}^j \d_{i_{l+2}}^k)\Big]
     \nn
 \eea
 %%%
 Using
 %%%
 \bea
 f^{a'fg}f^{a'bc}=- f^{a'gc}f^{a'bf}- f^{a'gb}f^{a'fc}
 \eea
 %%%
 one gets
 %%%
 \bea
&& 2(2F_1+F_2+F_3)\;N\d^{fg}\Big[
 f^{a'gc}f^{a'fb}(-2\d_{i_{l+2}}^j\d_{i_{l+1}}^k
        +\d_{i_{l+1}}^j\d_{i_{l+2}}^k+\d^{jk}\d_{i_{l+1}i_{l+2}})\nn\\
 &&\hspace*{1.2in}+f^{a'gb}f^{a'fc} (-2\d_{i_{l+2}}^k\d_{i_{l+1}}^j
        +\d_{i_{l+1}}^k\d_{i_{l+2}}^j+\d^{jk}\d_{i_{l+1}i_{l+2}})
   \Big]
 \eea
 %%%
 Due the presence of $\psi$ tensor (which enforces cyclicity) in \rf{VLna},
 this can be rewritten as
 %%%in
 \bea
&& -4(2F_1+F_2+F_3)_{div} N^2 \d_{bc}\Big[
(-2\d_{i_{l+2}}^j\d_{i_{l+1}}^k
        +\d_{i_{l+1}}^j\d_{i_{l+2}}^k+\d^{jk}\d_{i_{l+1}i_{l+2}})
   \Big]
    \label{xxxxfinal}
 \eea
 %%%
where the subscript in $(2F_1+F_2+F_3)_{div}$ indicates that only the divergent part containing the usual $\G(1-\w)$ factor has been kept.
\ni With the divergence part kept, eq.\rf{xxfinal} and eq.\rf{xxxxfinal} must be added according to \rf{renctr-nonabelian} for the complete result. Since our proposal does not fix the numerical relative numerical coefficients in \rf{vqmother_rel2q}, they can be adjusted in such a way that the SO(6) index factor of the sum takes the form of
 %%%
 \bea
 \d^{jk}\d_{i_{l+1}i_{l+2}}
        +2\d_{i_{l+1}}^j\d_{i_{l+2}}^k -2\d_{i_{l+2}}^j\d_{i_{l+1}}^k
 \eea
 %%%
which is the same as the SYM result in
\cite{Minahan:2002ve}.\footnote{The overall coefficient includes
factors of $\fr{1}{r_0}$ unlike the SYM result. This may not be an
essential difference. From the viewpoint of an open string, the
non-linear sigma in the curved background includes non-pertubative
effects as well whereas the classical SYM does not. Once the
non-perturbative effects are included, however, the resulting SYM
effective action becomes to contain terms of inverse powers in
$\phi^m\phi^m$ (see \cite{GonzalezRey:1998uh} for example). If one
performs perturbative analysis (which is only possible for abelian
case since the non-abelian version is not known) with such an
action, one will have to make a large $\phi_0$ expansion, therefore,
introducing a parameter $\phi_0$ (which would correspond to $r_0$).
Introduction of $\phi_0$ or $r_0$ will spontaneously break the
symmetry with non-linear realization of the symmetry. We are not
aware of any extensive discussion on how such non-linear realization
would affect amplitude computations. In the current case at least,
the effect seems moderate. On the SYM side, it will be an interesting task to explicitly check whether a non-abelian version of the action obtained in \cite{GonzalezRey:1998uh} would lead to the same result as obtained by Minahan and Zarembo up to a possible overall factor.} In a more systematic method that was mentioned previously, it should be possible to determine the numerical coefficients precisely. We expect that the coefficients so determined will be consistent with the present result.

\ni Below \rf{renctr-nonabelian}, we mentioned that the appearance of the extra factor $N$ (as compared with the abelian result \rf{renctr}) was deduced on general grounds. It is basically determined by the normalization of the group generators.
The non-abelianiztion of the cubic vertex in \rf{bctr} is more subtle for the following reason.   As stated a few times before, we expect that the non-abelianization that we have proposed should have its origin in a (yet unknown) formulation of the non-linear sigma model using some generalized algebra. For the diagram in Fig. 3, we need two factors of the cubic vertex,
 %%%
 \bea
 \int \;
     \pa X^m \pa X^m\; \fr{X_0\cdot X}{r_0^2}
      \pa X^m \pa X^m\; \fr{X_0\cdot X}{r_0^2}
                \label{cubicv}
 \eea
 %%%
The correct non-abelianization may not be an ordinary square of non-abelianization of a single cubic vertex because now the multiplication is that of the generalized algebra. Since there are more fields, it is harder to make a prediction the same way as for the quadratic and the quartic vertices.

%%%%%%%%%%%%%%%%%%%%%%%%%%%%%%%%%%%%%%%%%%%%%%%%%%%%%%%%%%%%
\subsection*{$L \geq 3$ cases }

The $L=2$ case captures most of the essential features of the analysis with a minimal amount of algebra. Nevertheless, there are a few aspects that appear only in $L\geq3$ cases.
In this subsection, we analyse $L \geq 3$ cases starting with $L=3$ case and comment on $L\geq 4$ cases afterwards. A more comprehensive account will be given in \cite{ipark}.

\ni For $L=3$ case, computation with the quadratic vertex is very similar: it gets contracted with one of the elementary states. The analysis of the quartic vertex reveals a feature that should be shared by the cases of arbitrary
$L$'s. The correlator to compute is
 %%%
 \bea
&&\int dy  <
 [X'^{\,i}(x_1)\otimes{a}][X'^{\,j}(x_2)\otimes{b}] [X'^{k}(x_3)\otimes{c}]\nn\\
 &&\quad\quad\quad [X'^{i_{l}}(x)\otimes{e}]  [X'^{i_{l+1}}(x)\otimes{f}] [X'^{i_{l+2}}(x)\otimes{g}]
       \nn\\
 && f_{a'b'c'}f_{a'd'e'}[X'^{\,i'}(y)\otimes{b'}] [X^{j'}(y)\otimes{c'}] [X'^{\,i'}(y)\otimes{d'}] [X^{j'}(y)\otimes{e'}]> %\label{L3ctr4leg1}
 \eea
 %%%
There are a few different types of terms. When there is a contraction among the fields within the first line, the resulting diagram is a disconnected one, therefore need not be considered.
The second type of terms are the ones that have no contraction within the first line but has one contraction between the first and second line. That particular contraction becomes a spectator line with the other part of the diagram becomes essentially the one-loop diagram of $L=2$.
We have explicitly confirmed this using Mathematica: $L=3$ case can be viewed as $L=2$ case with  the third leg standing as a spectator. It is rather evident that these kinds of grouping will remain valid for an arbitrary value of $L$.

%%%%%%%%%%%%%%%%%%%%%%%%%%%%%%%%%
\section{Conclusion}
%%%%%%%%%%%%%%%%%%%%%%%%%%%%%%%%%

\ni The focus of the previous works \cite{Park:2008fp}\cite{Park:2009ki} was scattering of {\em elementary} open string states. It was anticipated that the generation of the curvature would be revealed through a master counter vertex (i.e., pre $r_0$-expanded form of the counter vertex) that needed to be introduced to cancel the spacetime divergences. There are two things that make this task non-trivial. First, it seems to require three-loop computation to prove the generation of the curvature by massless scattering. (For massive, it may be seen at two-loop.) The other complication come from the complexity of the vertex that realizes a closed string state. The counter vertex operator has been interpreted to be a collective representation of
(infinitely many) closed strings. Undoubtedly it takes a complicated form, and can be viewed as an example of the long-known difficult task of the realization of a closed string in terms of an open string field.

\ni On retrospect of \cite{Gubser:2002tv}, the task is complicated partially because the closed string to be realized would represent a {\em "fundamental"} string. The works of \cite{Gubser:2002tv} and \cite{Minahan:2002ve}\cite{Tseytlin:2003ac} offer a new insight on the task. The task becomes much more tractable if one tries to realize a "solitonic" closed string configuration instead. In \cite{Gubser:2002tv} and \cite{Minahan:2002ve}, the authors considered SYM composite operators
in the context of AdS/CFT. In the present work, we have used an
open string setup for a related purpose.

In section 4, we carried out renormalization of $V_L$ given in \rf{VLna},
 %%%
 \bea
V_{L}\equiv  \psi_{m_1...m_L}\,[{X'}^{m_1}\otimes a_1]\cdots [{X'}^{m_L}\otimes a_L]\;
   \Tr(\l^{a_1}\cdots \l^{a_L})
 \label{VLnaqq}
 \eea
 %%%
Some of the statements that we have made in section 4 depend on its form, especially on
the absence of the fermionic coordinates in \rf{VLnaqq}. However,
it is expected that the full form of the vertex would contain the fermionic coordinates. Therefore, there remains two important tasks: obtaining the
fermionic parts and ensuring that they do not affect the final form of the result obtained in the previous section. Once the full vertex is obtained,
the second task will be straightforward even though it will require a vast amount
of algebra.
Let us imagine carrying out the first task in the Green-Schwarz formulation. In the Green-Schwarz formulation, one would utilize the correspondence between the SYM scalar, $\phi^m$, and the following vertex operator,
 %%%
 \bea
 \phi^m\quad \Longleftrightarrow
    \quad V_s\equiv ({X'}^{m}+R^{mv}k^{v})\;e^{ik\cdot X}
    \label{scalarvertex}
 \eea
 %%%
 where the index $v$ runs along the worldvolume directions.
Naively,\footnote{We will shortly argue that \rf{VLnaive} may
not be the correct vertex operator. Even if it (or its
modification) is the correct vertex, considering \rf{VLnaqq} can be
justified for the following practical reasons.
 Firstly,
\rf{VLnaive} reduces to \rf{VL} by setting $k=0$; the computations that involve \rf{VL} will be part of those that involve \rf{VLnaive} or its possible modification. Secondly, \rf{VL} is the bosonic part
of \rf{VLnaive} other than the exponential factor. As observed in
\cite{Park:2008fp} and \cite{Park:2009ki}, the fermionic terms have a
strong tendency to lead to vanishing results. It might be that
they do not make additional contributions anyway.
}
this seems to suggest that the corresponding open string vertex operator might be
 %%%
 \bea
V_{L,k}\equiv \psi_{m_1...
m_L}\Big[ ({X'}^{m_1}+R^{m_1v_1}k^{v_1})\cdots ({X'}^{m_L}+R^{m_Lv_L}k^{v_L}) \Big]\; e^{ik\cdot X}
 \label{VLnaive}
 \eea
 %%%
There is an aspect of the remaining procedure that differentiates the construction of a would-be closed string operator from that of an ordinary open string, massless or massive. To better illustrate the point, we remind of what is commonly done
 for an ordinary open string in the pure spinor formulation.
 In the pure spinor approach, a preliminary form
of a vertex is written down in terms of the relevant superfields of certain
conformal dimensions. The free field equations for the superfields are known. (They are known for the first few levels of the open string spectrum, and in principle it should be possible to obtain the field equations for higher levels as well.) The vertex becomes on-shell after requiring it to vanish by the BRST charge, and the field equations are reproduced.
However, if the state to be represented is a soliton of the two-dimensional non-linear sigma model on, say, AdS$_5\times$S$^5$, what is its field
equation?

This seems to imply that a different approach is required in construction of a vertex operator that represents a closed string. For the geometric vertex operator $V_G$, it was the geometry of the D-brane that provided the guide.
For $V_L$, the only guide that is available for its (preliminary) form may be the group/symmetry properties. Interestingly, this is a situation similar to
when one constructs the SYM operator \rf{o-quoted}. Another implication that is not unrelated to the first seems to be that one does not have means to determine the on-shell value of $k^v$ that appears in the pure spinor analogue of \rf{VLnaive}. It is less directly obvious in the Green-Schwarz formulation where the
on-shell momentum follows from requiring conformal invariance.
In other words, one imposes the physical state
condition on an ordinary open string state to ensure essentially the conformal gauge choice independence of amplitude computations.
Again it may be useful to recall what was done (or not done) for $V_G$. For $V_G$, we did not impose such a condition in our previous works, therefore, the conformal symmetry issue may need to be checked.\footnote{We
thank P. Ho for pointing this out.} As a matter of fact, the appearance of the closed string(s) in
Fig.1 signifies breaking of the 2D conformal symmetry,\footnote{that is analogous to spontaneous symmetry breaking in QFT } and what we have been up to should correspond to renormalization of $V_L$ in the 2D theory with
spontaneously broken conformal symmetry.

 Although conformal symmetry is highly
useful in the first quantization, it is not essential in (a certain version of) string field theory (see, e.g., \cite{hep-th/0311264}).
Presumably what is needed is various consistency checks on \rf{VLnaqq}
and a systematic study of conformal symmetry breaking in the first
quantization.

Another issue in need of further study is the way that \rf{quartic vertices} has been introduced: strictly speaking it is a choice. It remains to be seen whether one can better motivate or even achieve the first principle derivation of \rf{quartic vertices}. This is a direction that we are actively pursuing currently, and will report on the progress in the near future. The non-abelianization of the internal vertices must have deeper origin.
We believe that a proper formulation of the non-linear action for an open string in a multi-brane background would require use of a generalized Myer's effect \cite{Myers:1999ps} and a generalized algebra. (Presumably this issue is related to the recent development in the multiple M2-brane \cite{Bagger:2007jr}. We find the work of \cite{Park:2008qe} interesting in this regard.) In turn, the action thereby constructed could be mapped to an action where the group generators replace the elements of the generalized algebra. Such mapping should be possible as explored in, i.e., in \cite{Gustavsson:2010ep} in the context of the three-algebra. Once the elements of the generalized algebra are mapped to matrices, there will not be many choices (other than the way that has been proposed in this work) in contracting the group indices in \rf{quartic indices} since typically the action requires a symmetrized trace. Constructing a lower dimensional non-abelian string/brane action starting with a higher dimensional one, therefore, will inevitably involve "states selection" that is associated with dimensional reduction and truncation. To a large extent, the choice would be dictated by or attributed to the physics under consideration. In our case, the physics under consideration is the inclusion of the vertex operator in \rf{o-quoted}

We end with a few more remarks on the implications of our result for AdS/CFT. It is clear that the "small $\a'$" limit is essential to reduce open string theory to the SYM theory. Put differently, if one does not take that limit, there will be open stringy contributions that cannot be neglected. Nevertheless, there will still be a duality that is a generalized version of AdS/CFT. The $AdS$ geometry would still result, since, in essence, it comes from the S-duality. Therefore, it may well be that one has a $AdS$ closed string on one side whereas the dual side is open string obtained by taking some convenient/simplifying limit that is not precisely the limit given in \rf{sym limit}. The second remark is with regard to the roles the system branes and a virtual brane in Fig. 1. The picture seems to provide a logical rationale (although at a pictorial level) for existence of AdS/CFT and its generalization. One interesting aspect of Fig. 1 is as follows. What would be the meaning of interchanging the roles of the system branes and the virtual brane? The figure as it stands is suitable for a setup where one starts with an open string and tries to realize a closed string. Interchanging the roles of the system branes and the source branes would, therefore, correspond to setup where one starts with a closed string and tries to realize an open string degrees of freedom in terms of the closed string fields. That, presumably, will require spontaneous symmetry breaking.  At a quantitative level, the first step would be along the line of \cite{Sato:2003ky} where it was shown that there are a class of supergravity solutions that can be put into a form of a DBI action.

\newpage
%%%%%%%%%%%%%%%%%%%%%%%%%%%%%%%%%%%%

\end{document}